# Functionalized nanopore-embedded electrodes for rapid DNA sequencing

*Haiying He, Ralph H. Scheicher[*], and Ravindra Pandey[*]*
Department of Physics and Multi-Scale Technologies Institute,
Michigan Technological University, Houghton, MI 49931, U.S.A.

*Alexandre Reily Rocha[†] and Stefano Sanvito*
School of Physics and CRANN, Trinity College, Dublin 2, Ireland

*Anton Grigoriev, Rajeev Ahuja[‡]*
Condensed Matter Theory Group, Department of Physics, Box 530,
Uppsala University, S-75121 Uppsala, Sweden

*Shashi P. Karna*
US Army Research Laboratory, Weapons and Materials Research Directorate,
ATTN: AMSRD-ARL-WM, Aberdeen Proving Ground, MD 21005-5069, U.S.A.

[†] *Present address:* Department of Materials Physics and Mechanics, University of São Paulo, São Paulo, Brazil.
[‡] *Also at:* Applied Materials Physics, Department of Materials and Engineering, Royal Institute of Technology (KTH), S-00 44, Stockholm, Sweden

[*]*Corresponding Authors:*
Ralph H. Scheicher: rhs@mtu.edu
Ravindra Pandey: pandey@mtu.edu
Tel: +1 (906) 487-2086, Fax: +1 (906) 487-2933

**The determination of a patient's DNA sequence can, in principle, reveal an increased risk to fall ill with particular diseases [1,2] and help to design "personalized medicine" [3]. Moreover, statistical studies and comparison of genomes [4] of a large number of individuals are crucial for the analysis of mutations [5] and hereditary diseases, paving the way to preventive medicine [6]. DNA sequencing is, however, currently still a vastly time-consuming and very expensive task [4], consisting of pre-processing steps, the actual sequencing using the Sanger method, and post-processing in the form of data analysis [7]. Here we propose a new approach that relies on functionalized nanopore-embedded electrodes to achieve an unambiguous distinction of the four nucleic acid bases in the DNA sequencing process. This represents a significant improvement over previously studied designs [8,9] which cannot reliably distinguish all four bases of DNA. The transport properties of the setup investigated by us, employing state-of-the-art density functional theory together with the non-equilibrium Green's Function method, leads to current responses that differ by at least one order of magnitude for different bases and can thus provide a much more robust read-out of the base sequence. The implementation of our proposed setup could thus lead to a viable protocol for rapid DNA sequencing with significant consequences for the future of genome related research in particular and health care in general.**

The possibility of passing a DNA strand through a so-called "nanopore" with a diameter of only a few nanometers has been explored [10-14] for the purpose of DNA sequencing. A negatively charged single-stranded DNA (ssDNA) molecule (in solution with counter ions) is driven by an electric field from one side of a membrane to the other through the nanopore. As the nucleotides of the DNA are migrated across the membrane, they will partially block the pore in different ways depending on their size [15]. Thus, the ionic current through the pore due to passing counter ions is characteristically affected, and monitoring the ionic blockade [13,16] could lead to the determination of the DNA sequence. To resolve the remaining ambiguity between the pyrimidine bases cytosine (C) and thymine (T), and the purine bases adenine (A) and guanine (G), it was recently suggested [17] that one could embed electrodes in the walls of a solid-state nanopore [18]. By applying a bias voltage across the electrodes a small electric current perpendicular to the DNA strand can be measured. Thus as the ssDNA migrates through the pore, time-dependent current-voltage signals from the electrodes would supposedly be specific enough to allow the unequivocal identification of the nucleobase passing across the pore at a given moment. The resulting time sequence of the current-voltage signal could then directly be translated into the corresponding DNA sequence. It is, however, still unclear whether or not the resulting current-voltage signal from pore-embedded electrodes is sensitive enough to reliably distinguish the four nucleic acid bases of DNA. The signals for different nucleotides are found to be identical within statistical error [8,9]. Although it may be possible to resolve the nucleic acid bases using statistical analysis [9], this approach requires previous calibration via repeated sampling of a known sequence, and the resulting signal currents are all of the same order of magnitude. For a specific identification, it would be desirable to have electrical current response for each nucleobase, which differ by one order of magnitude or more. Here we present the results of a systematic *first-principles* study to test the feasibility of introducing functionalized electrodes for nanopore DNA-sequencing.

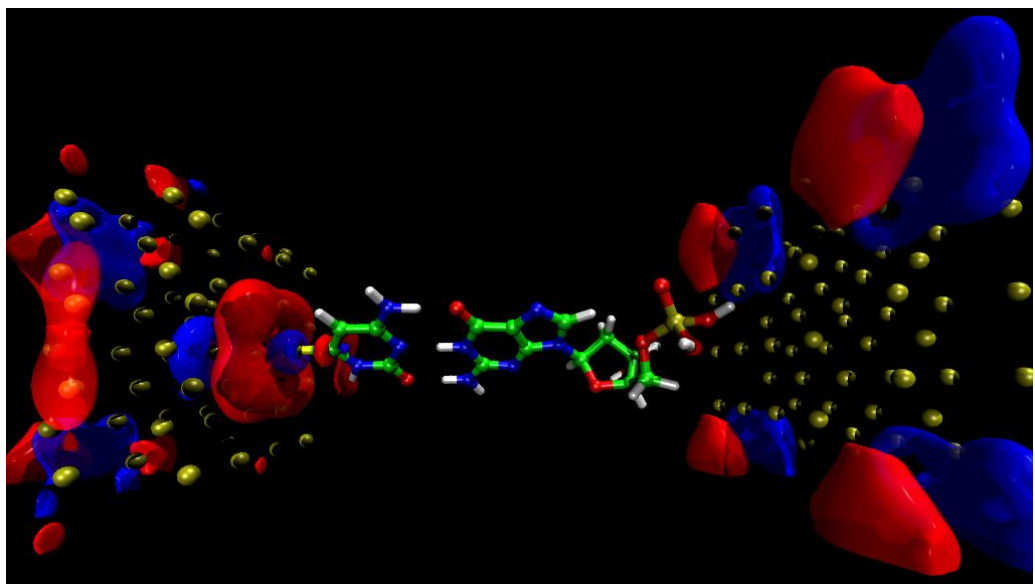

**Fig. 1: An illustration of the proposed device: ssDNA is passing through the nanopore with functionalized gold electrodes embedded. A probe molecule (e.g. Cytosine) is immobilized on the inner surface of the left electrode by a sulfur atom. As part of the ssDNA molecule, a Guanosine monophosphate unit is shown as the target to be identified. The wave function of the first occupied contact state corresponding to the transmission peak labeled by * in Fig. 4 is also shown, with the colors blue and red indicating the phase.**

A schematic illustration of the key elements of the proposed device is shown in Fig. 1. It consists of two metallic gold electrodes embedded opposite to each other in a nanopore through which a ssDNA molecule can pass. A chosen purine or pyrimidine DNA base molecule which acts as the molecular probe of the functionalized electrode is anchored to the inner surface of one of the electrodes via a thiol group. As a ssDNA is pulled through the nanopore via an applied electric field parallel to the nanopore walls, the probing molecule will simultaneously fulfill two functions: (i) stabilization of the target base in the DNA sequence by forming weak hydrogen bonds, and (ii) detection of the target base by coupling electronically to it. Since the nucleic acid bases occurring in natural DNA possess an inherent ability to selectively bind to their respective complementary base partners, all four base molecules (i.e. Adenine (A), Cytosine (C), Guanine (G), and Thymine (T)) are considered as the probe molecules in calculations.

We have started our investigation by performing a series of electronic structure calculations to determine the most stable pairing geometries between the probe and the target molecule. For each pair, five initial positions were considered corresponding to the five matching position at the H-bonding edge. We employed density functional theory (DFT) within the generalized gradient approximation (GGA) of the exchange and correlation functional, as parameterized by Perdew, Burke and Ernzerhof (PBE) [19], incorporated in SIESTA [20]. The bias-dependent tunneling current is then calculated from the non-equilibrium Green's Function method based on the Keldysh formalism, as implemented in SMEAGOL [21-23]. The binding energies ($E_b$) of the "probe-target" base pairs formed temporarily in the nanopore are strongly correlated to the number of H-bonds formed as well as to the steric/electrostatic repulsion of atoms having similar electron affinities (see Table I). The calculated values for $E_b$ show that the H-bonding can significantly stabilize the DNA molecule as it passes through the pore, thereby preventing drastic variations in the current due to thermal fluctuations of the structure and the uncertainty in the orientation of the base between the two electrodes. On the other hand, the H-bonding between probe- and target-base is weak enough to allow for an easy break-up when the ssDNA molecule is pulled through the nanopore by the electric field. The covalent bond of the probe-base's thiol group is, however, strong enough to keep the probe firmly placed on the surface of the gold electrode, without the risk that it would be swept along with the passing DNA molecule.

**Table I: Summary of the information deducible for different probes from current measurements (at 100 mV and 250 mV) or from a force signal (proportional to the binding energy). Target bases that cannot be distinguished are combined in parentheses.**

| Probe\Bias | V=100 mV | V=250 mV | Binding energy ($E_b$) |
|---|---|---|---|
| A | (A, C, G) from T | A from (C, G) from T | (A, C) from (G, T) |
| C | (A, G) from (C, T) | A from G, C from T | (A, C) from G from T |
| G | (A, C, G) from T | (A, C, G) from T | (A, T) from C from G |
| T | (A, C, G) from T | C from (A, G) from T | (A, C, G, T) |

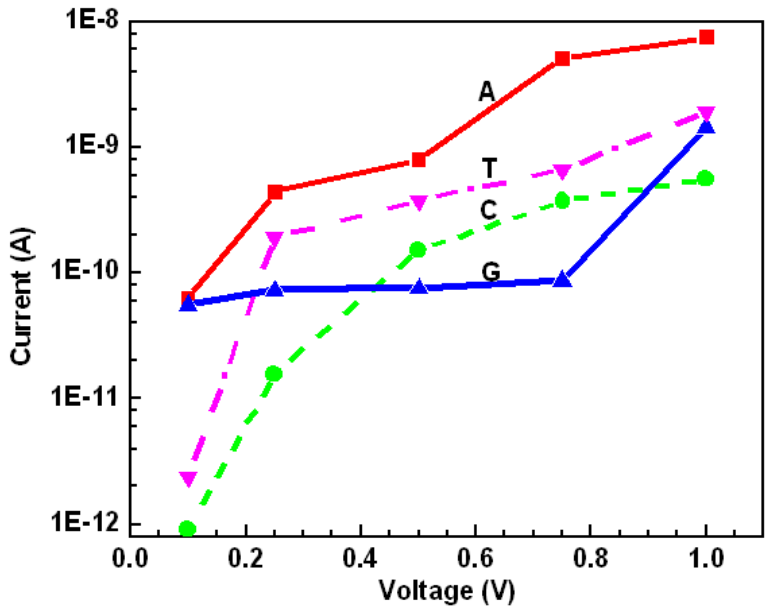

**Fig. 2: The current-voltage curve for the device functionalized with a C-probe for all four possible target bases (A: red square; C: green circle; G: blue triangle; T: pink upside-down triangle) with current signals plotted in logarithmic scale.**

The current-voltage curves calculated for the device involving the nucleobase C as probe are shown in Fig. 2. The gap between target base and gold electrode (Fig. 1) leads to relatively small current values at zero bias. Recalling that the figure of merit for distinguishing two different base molecules is that their associated currents differ by at least one order of magnitude, in Table I we present our recognition map. For instance at a bias of 100 mV, using either A, G, or T as a probe, we can distinguish the set A, C, or G from T (while A, C and G cannot be distinguished from each other). In contrast, when C is used as a probe, we can differentiate A and G from C and T. At an increased bias the recognition properties of the proposed device changes. For instance at 250 mV, the A probe provides different currents for A, T and for the C, G pair (C and G however remain indistinguishable from each other). At the same voltage the C probe can distinguish A from G, and C from T, the G probe can distinguish the set (A, G, C) from T, and finally the T probe can distinguish C from A or G, and from T. It therefore appears that T is the most easily to identify base, as also suggested in a previous study [8]. Our scheme, however, adds the ability of identifying A and G, which used to be the major obstacle of nanopore DNA sequencing. In our calculations, this is achievable by using A or C as probe at a bias of 250 mV.

Functionalization of the electrodes with C as a probe appears to yield the best way for a reliable identification of all four base molecules in DNA. This can be achieved by carrying out three sequencing runs at three different bias voltages, namely 100 mV, 250 mV, and 750 mV. The flow diagram shown in Fig. 3 gives the illustration of this proposed protocol of DNA sequencing.

The first set of measurements at 100 mV would result in a series of current signals that fall into two easily distinguishable categories: "high" current values if A or G is the target base, and "low" current values if C or T is the target base. The difference between the two categories is nearly two orders of magnitude, which should make the distinction extraordinarily robust. We now require additional information to resolve the remaining ambiguity between A/G and C/T. In a second measurement at 250 mV, it will be possible to distinguish between C and T, as their respective current values differ by one order of magnitude at that bias voltage. Thus, any "high" current value would lead to the identification of a C in the sequence, while any "low" current value means that a T is at this position in the sequence. Finally, a third measurement at 750 mV causes the current values for the bases A and G to differ by two orders of magnitude, leading to an easy distinction between the two, where "high" current values correspond to A, while "low" current values correspond to G.

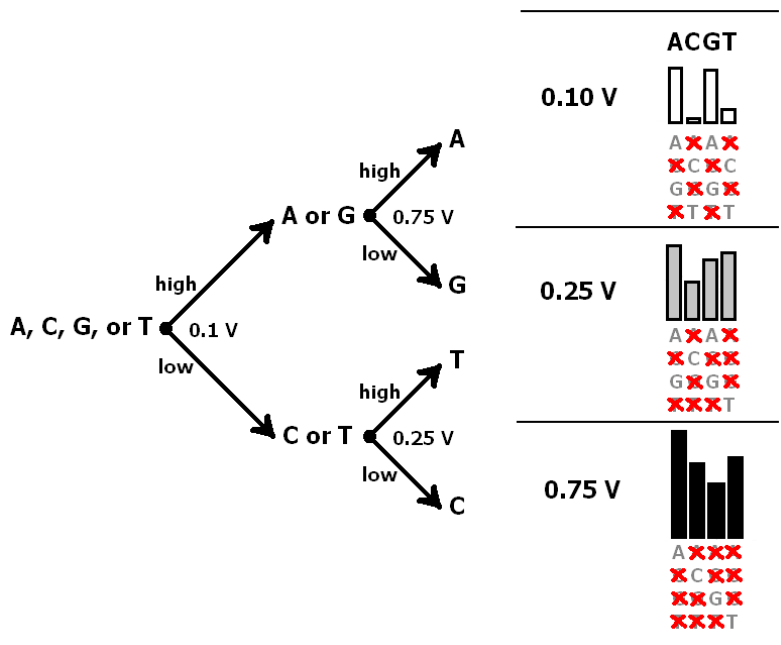

**Fig. 3: Flow diagram illustrating the decision-making process of a device involving C-probe leading to the identification of a target base in the sequence. Here, "high" and "low" refer to higher or lower current values at a given bias voltage. The height of the bars below the letters A, C, G, and T on the right side of the figure corresponds to the respective current signal (on a logarithmic scale). The crossed-out letters below the bars refer to possible target bases that have been ruled out.**

Having demonstrated the principal capability of our hypothetical device for reliable DNA sequencing, we now turn our attention to understand the electrical response for different target nucleic acid bases. In the following discussion, we concentrate on the most promising setup involving the C probe. The tunneling current (Fig. 2) is obtained by integrating the transmission function shown in Fig. 4 for each target base predominantly within the voltage window [23, 24]. For the given nanopore (2.18 nm wide; see supplementary information), the similarity in the zero-bias transmission-function shape for all target nucleotides can be attributed to the fact that the peaks near $E_F$ and within the molecular HOMO-LUMO gap, away from specific molecular orbitals, are mostly associated with the contact states localized on the anchoring group and the gold electrodes (Fig. 1). The size of the target base, thereby the gap between target and the (right) gold electrode determines the transmission magnitude via the overlap between the electronic states localized on the molecule and the states on the right electrode. Nevertheless, we would like to emphasize here that it is the different effect of the target base on the bias-dependent shift of these peaks that allows one to distinguish between the four nucleic acid bases.

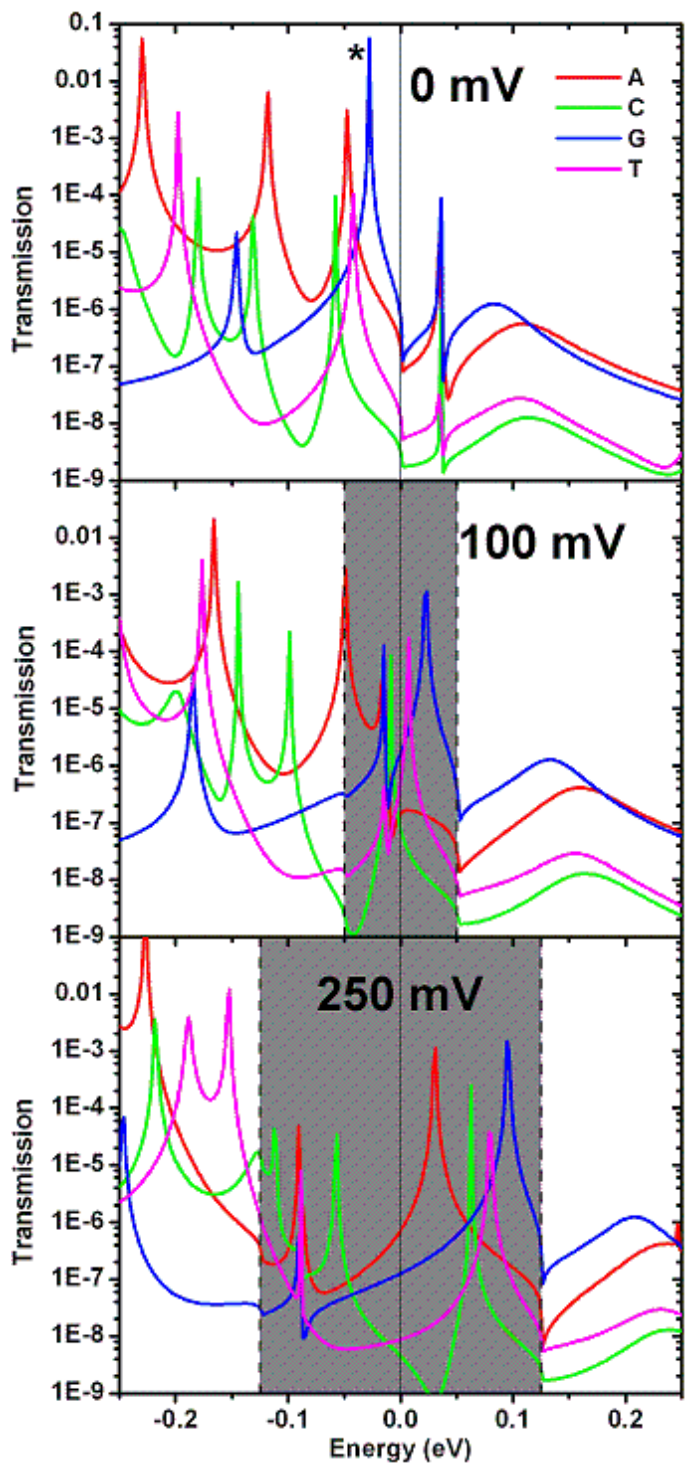

**Fig. 4: Device with C-Probe: Transmission functions for four target bases at a bias of V = 0 mV, 100 mV, and 250 mV with transmission values plotted in logarithmic scale. The Fermi level is aligned to zero. The bias window of ±V/2 is indicated by the shaded area. The transmission peak for target "G" at zero bias corresponding to the molecular orbital shown in Fig. 1 is labeled by \*.**

When the bias is increased from zero to 250 mV, the transmission peaks associated with the contact states enter the voltage window and contribute to the increase of the current across the device (e.g., the potential on the left electrode increases, thus causing a rise in the energy of the peaks). The position of the peak related to the contact state closely follows the shift of the electrode potential and is only slightly affected by charging/discharging of the state when it is driven between the occupied state on the left electrode and the unoccupied on the right.

It may even be possible in our proposed setup to combine the electrical signal from the transverse current with a force signal [25] driving the ssDNA through the nanopore modified with an anchored probe molecule. Since the binding energy of the selected probe to the target through the H bond is sequence specific, a different force needs to be applied to ensure the DNA strand traverses at a constant speed through the nanopore (see more in supplementary materials). This may provide supplemental information, in addition to the tunneling current, for distinguishing the DNA bases. For instance, using C as the probe, the current results at 100 mV allow for a direct distinction between the sets (A, G) and (C, T), while the binding energies can tell the difference between A and G, and between C and T (Table I). By combining the information of the driving force (threshold voltage) with the current results, it would suffice to have a single run with one probe at a given bias to distinguish all four nucleic acid bases.

In summary, we have shown that nanopore-embedded gold electrodes functionalized with molecular probes can lead to a dramatic improvement in the sensitivity of base molecules sensing in a pore-translocating DNA sequence. Our proposed approach focusing on functionalized electrodes for rapid nanopore DNA-sequencing could thus make this method feasible. If the described setup could be successfully implemented, then our findings might have a significant impact as a rapid and most definitive diagnostic tool.

*Acknowledgment*

Helpful discussions with S. Gowtham and K. C. Lau are acknowledged. The work at Michigan Technological University was supported by DARPA (contract number ARL-DAAD17-03-C-0115). AG and RA gratefully acknowledge financial support from Carl Tryggers Stiftelse för Vetenskaplig Forskning. ARR and SS thank Science Foundation of Ireland (grants SFI02/IN1/I175 and SFI05/RFP/PHY0062) for financial support.

# References

1. Futreal, P. A. *et al.* Cancer and genomics. *Nature* **409**, 850-852 (2001).

2. Jimenez-Sanchez, G., Childs, B. & Valle, D. Human disease genes. *Nature* **409**, 853-855 (2001).

3. Chakravarti, A. Single nucleotide polymorphisms: . . .to a future of genetic medicine. *Nature* **409**, 822-823 (2001).

4. Church, G. M. Genomes for ALL. *Scientific American* **294**, 47-56 (2006).

5. Stoneking, M. Single nucleotide polymorphisms: From the evolutionary past. . .. *Nature* **409**, 821-822 (2001).

6. Hood, L., Heath, J.P., Phelps, M.E. & Lin, B. Systems biology and new technologies enable predictive and preventative medicine. *Science* **306**, 640-643 (2004).

7. International Human Genome Sequencing Consortium, Initial sequencing and analysis of the human

genome. *Nature* **409**, 860-921 (2001).

8. Zwolak, M. & Di Ventra, M. Electronic signature of DNA nucleotides via transverse transport. *Nano Lett.* **5**, 421-424 (2005).
9. Lagerqvist, J., Zwolak, M. & Di Ventra, M. Fast DNA sequencing via transverse electronic transport. *Nano Lett.* **6**, 779-782 (2006).
10. Kasianowicz, J. J., Brandin, E., Branton, D. & Deamer, D. W. Characterization of individual polynucleotide molecules using a membrane channel. *Proc. Natl. Acad. Sci. USA* **93**, 13770-13773 (1996).
11. Akeson, M., Branton, D., Kasianowicz, J. J., Brandin, E. & Deamer, D. W. Microsecond time-scale discrimination among polycytidylic acid, polyadenylic acid, and polyuridylic acid as homopolymers or as segments within single RNA molecules. *Biophysical Journal* **77**, 3227-3233 (1999).
12. Meller, A., Nivon, L., Brandin, E., Golovchenko, J. & Branton, D. Rapid nanopore discrimination between single polynucleotide molecules. *Proc. Natl. Acad. Sci. USA* **97**, 1079-1084 (2000).
13. Deamer, D. W. & Akeson, M. Nanopores and nucleic acids: prospects for ultrarapid sequencing. *Trends Biotechnol.* **18**, 147-151 (2000) and references therein.
14. Khanna, V. K. Existing and emerging detection technologies for DNA (Deoxyribonucleic Acid) finger printing, sequencing, bio- and analytical chips: A multidisciplinary development unifying molecular biology, chemical and electronics engineering. *Biotechnology Advances* **25**, 85-98 (2007).
15. Fologea, D. *et al.* Detecting single stranded DNA with a solid state nanopore. *Nano Lett.* **5**, 1905-1909 (2005).
16. Vercoutere, W. A. *et al.* Discrimination among individual Watson-Crick base pairs at the termini of single DNA hairpin molecules. *Nucleic Acids Research* **31**, 1311-1318 (2003).
17. Li, J., Gershow, M., Stein, D., Brandin, E. & Golovchenko, J. A. DNA molecules and configurations in a solidstate nanopore microscope. *Nature Materials* **2**, 611-615 (2003).
18. Rhee, M. & Burns, M. A. Nanopore sequencing technology: nanopore preparations. *Trends Biotechnol.* **25**, 174-181 (2007).
19. Perdew, J. P., Burke, K. & Ernzerhof, M. Generalized gradient approximation made simple. *Phys. Rev. Lett.* **77**, 3865-3868 (1996).
20. Soler, J. M. *et al.* The SIESTA method for ab initio order-N materials simulation. *J. Phys. Condens. Matter* **14**, 2745-2780 (2002).
21. Rocha, A. R. *et al.* Computer code SMEAGOL (Spin and Molecular Electronics in an Atomically Generated Orbital Landscape), www.smeagol.tcd.ie.
22. Rocha, A. R. *et al.* Towards molecular spintronics. *Nat. Mater.* **4**, 335-339 (2005).
23. Rocha, A. R. *et al.* Spin and molecular electronics in atomically generated orbital landscapes. *Phys. Rev. B* **73**, 085414-1-085414-22, (2006).
24. Grigoriev, A. and Ahuja, R. Molecular Electronics Devices in Nano and Molecular Electronics Handbook, editor S.E. Lyshevski, CRC Press (2007).
25. Zhao, Q. *et al.* Detecting SNPs using a synthetic nanopore. *Nano Lett.* **7**, 1680-1685 (2007).

# Supplementary Information

**Computational Methods**

**Simulation model.** As illustrated in Fig. 1, the proposed device consists of two metallic gold electrodes embedded opposite from each other across the diameter of a nanopore, with a ssDNA molecule passing through it. A chosen purine or pyrimidine DNA base molecule is anchored to the inner surface of one of the electrodes via a thiol group, acting as the molecular probe of the functionalized electrode. As ssDNA is pulled through the nanopore, the probe molecule will simultaneously stabilize (by forming weak hydrogen bonds) and detect (by coupling electronically) one nucleobase in the ssDNA at a time.

We have investigated four possibilities for the probe base molecule based on the four nucleic acid bases occurring in natural DNA, namely Adenine (A), Cytosine (C), Guanine (G), and Thymine (T). The probe molecule is anchored by a sulfur atom at the hole site of Au(001) surface. The orientation of the probe molecule, however, is assumed to have the H-bonding edge parallel to the electrode surface. Due to the apparent difference in the size of the four bases, distance between the two electrodes is changed according to the size of the probe base ensuring the current measurements within the same detecting scale. For example, the distance between electrodes was 2.27 nm when the purine bases A or G were acting as the probe, and 2.18 nm in the case of the smaller pyrimidine bases C or T.

The molecular probe couples electronically only to a single nucleotide at a time. Therefore, it is sufficient to include only a fraction of the ssDNA molecule, namely one nucleotide (a DNA base terminated by the phosphate-sugar backbone) in the central scattering region of our transverse electron transport calculations. The contributions from neighboring bases are negligibly small in the setup chosen by us due to the large vacuum space between the electrode and the molecular part positioned off the probe. Depending upon the size of the target base, the vacuum length left between the right-side of the DNA molecule and the right electrode will be different.

**Electronic structure and geometry optimization.** The pairing geometry between the probe and the target molecule is obtained by carrying a full optimization of an isolated pair of two bases terminated by methyl groups. We used the parameterization of Perdew, Burke and Ernzerhof (PBE) [19] for the exchange-correlation functional, and norm-conserving pseudopotentials [26] with nonlinear core correction [27]. The employed reference configuration was $2s^2 2p^2 3d^0 4f^0$, $2s^2 2p^3 3d^0 4f^0$, $2s^2 2p^4 3d^0 4f^0$, $3s^2 3p^4 3d^0 4f^0$, $1s^1 2p^0 3d^0 4f^0$, for C, N, O, S and H respectively. Valence wave functions were expanded in a SIESTA localized basis set (numerical localized pseudo-atomic orbitals) [28]. The finite range of the orbitals was defined by an orbital confinement energy [29] of 50 meV. We used a double-zeta basis set with polarization orbitals on all the atoms [28]. The resolution of the real-space mesh was defined by a 150 Ry cutoff, assuring energy and force convergence. The tolerance in maximum density matrix difference is $10^{-5}$ and the tolerance in maximum atomic force is 0.04 eV/Å.

The optimized structures of base pairs are then used for transport calculation after substituting the methyl group with the phosphate-sugar backbone. The orientation as well as the structure of the phosphate-sugar is taken from the experimental structure of poly-nucleotides [30,31] and kept fixed for all the four nucleotides.

**NEGF quantum transport formalism.** Under the assumption of coherent transport, the electronic transport calculations for such a molecular system coupled to Au electrodes (Fig. 1) are calculated from the non-equilibrium Green's Function method based on the Keldysh formalism, encoded in SMEAGOL, which has been successfully applied to the study of electron transport in molecular systems. The semi-infinite electrodes act as charge reservoirs with electrons following the Fermi-Dirac distribution under a certain temperature (300 K in this case). A central scattering region is defined to include four 4×4 **R** Au layers along the (001) direction on either side of the electrodes in a periodic supercell as well as the DNA probe-target bases. This region is chosen to enclose the significant change in the electronic structure due to the presence of a bias. And the left and right electrodes outside this region are ensured to be not affected by the rearrangement within this region owing to the short screening length of Au [32]. The semi-infinite effect of the left (right) electrode is taken into account by introducing the self-energy $\Sigma_L$ ($\Sigma_R$) in the effective Hamiltonian [23]

$$G_{L(R)} C_{L(R)} \Sigma_{L(R)} = C^{+}_{L(R)} G_{L(R)} C_{L(R)}$$

$$G(E) = [ES - H_S - \Sigma_L(E) - \Sigma_R(E)]^{-1}$$

The total current in such a device is given by

$$I = \frac{2e}{h} \int_{-\infty}^{\infty} dE\, T(E,V)[f(E-\mu_1) - f(E-\mu_2)]$$

where $\mu_1$ and $\mu_2$ are the electrochemical potentials in the two contacts under an external bias $V$, $f(E)$ is the Fermi-Dirac distribution function. The prefactor value of two is used to take into account the spin degree of freedom.

The transmission function, $T(E,V)$ is an important intrinsic factor describing the quantum mechanical transmission probabilities for electrons.

$$T(E,V) = Tr[\Gamma_L(E) G(E,V) \Gamma_R(E) G^{+}(E,V)] \quad \Gamma_{L(R)}(E) = i(\Sigma_{L(R)} - \Sigma^{+}_{L(R)})$$

It is worth noting that the transmission probability depends not only on the electron energy $E$ but also upon the applied external bias $V$. More details of the theoretical scheme see Refs. 21-23.

In the current calculation, the positions of Au atoms in the electrodes are fixed at the bulk value of a face-centered cubic structure with a lattice constant of 4.08 Å. We have used a Monkhorst Pack grid of 4×4×20 in the Au lead calculation, which sets the boundary condition at the interface between the scattering region and the electrodes. The S anchoring atom lies at the hole site of Au(100) surface with Au-S bond lengths close to an average value of 2.40 Å. Similar to our structure optimization, the PBE exchange-correlation functional form and norm-conserving pseudopotentials were used in Smeagol transport calculations. But in considering the computational cost, single-zeta basis sets with polarization orbitals were used for C, N, O, S and H; while a single-zeta basis set was used for Au with only 6s as the valence electron. Note that the Fermi level in the bulk Au is largely dominated by a broad Au 6s band. The 5d and 6p orbitals only start to play a role at energies 3 eV below or 3 eV above the $E_F$ respectively. Since we are only interested in the low-bias (≤1.0 V) regime for transport calculations, the electrodes can be satisfactorily described by the 6s electron only. The resolution of the real-space mesh was again set to a 150 Ry cutoff, and the tolerance in maximum density matrix difference is $10^{-4}$. The charge density is obtained by integrating the Green's function over 200 imaginary and 1000 real energy points according to the scheme described in Ref. 23.

**Additional notes on results and discussion**

The results of tunneling currents corresponding to only one polarity of the bias voltage are given here. The conducting part of the diode-like rectifying I-V curve was used to distinguish the different base pairs of DNA.

Using a force signal to detect different DNA bases, a different force needs to be applied to ensure the DNA strand traverses at a constant speed through the nanopore. The applied force is linearly proportional to the binding energy of the probe-target pair. For a binding energy of 0.5 eV, the rupture of about 1 Å occurs with a force ~1 nN. Considering a typical thickness of 10 nm for a synthesized nanopore, the voltage required for such a force is about 2 V approximated by a uniform electric field across the nanopore. The external driving voltage is linearly scaled with the force, thereby the binding energy. The double/triple of the binding energy doubles/triples the voltage, which becomes detectable for pairs with different numbers of H bonds.

**Additional References to supplemental materials**

26. Troullier, N. and Martins, J. L. Efficient pseudopotentials for plane-wave calculations. *Phys. Rev. B* **43**, 1993-2006 (1991).

27. S. G. Louie, S. Froyen, and M. L. Cohen, Nonlinear ionic pseudopotentials in spin-density-functional calculations. *Phys. Rev. B* **26**, 1738-1742 (1982).

28. E. Artacho, D. Sánchez-Portal, P. Ordejón, A. García, and J. M. Soler, Linear-Scaling ab-initio Calculations for Large and Complex Systems. *Phys. Stat. Sol. B* **215**, 809-817 (1999).

29. J. M. Soler, E. Artacho, J. D. Gale, A. García, J. Junquera, P. Ordejóon, and D. Sá anchez-Portal, The SIESTA method for ab initio order-N materials simulation. *J. Phys.: Condens. Matter* **14**, 2745-2779 (2002).

30. Berman, H. M., W. K. Olson, D. L. Beveridge, J. Westbrook, A. Gelbin, T. Demeny, S. H. Hsieh, A. R. Srinivasan, and B. Schneider, The nucleic acid database. A comprehensive relational database of three-dimensional structures of nucleic acids. *Biophys. J.* **63**, 751-759 (1992).

31. Tjandra, N., Tate, S., Ono, A., Kainosho, M. , and Bax, A. The NMR structure of a DNA dodecamer in an aqueous dilute liquid crystalline phase. *J. Am. Chem. Soc*. **122**, 6190-6200 (2000).

32. Pettifor, D. *Bonding and Structure of Molecules and Solids*, Oxford University Press, Oxford, 2002.

**Supplementary figure:**

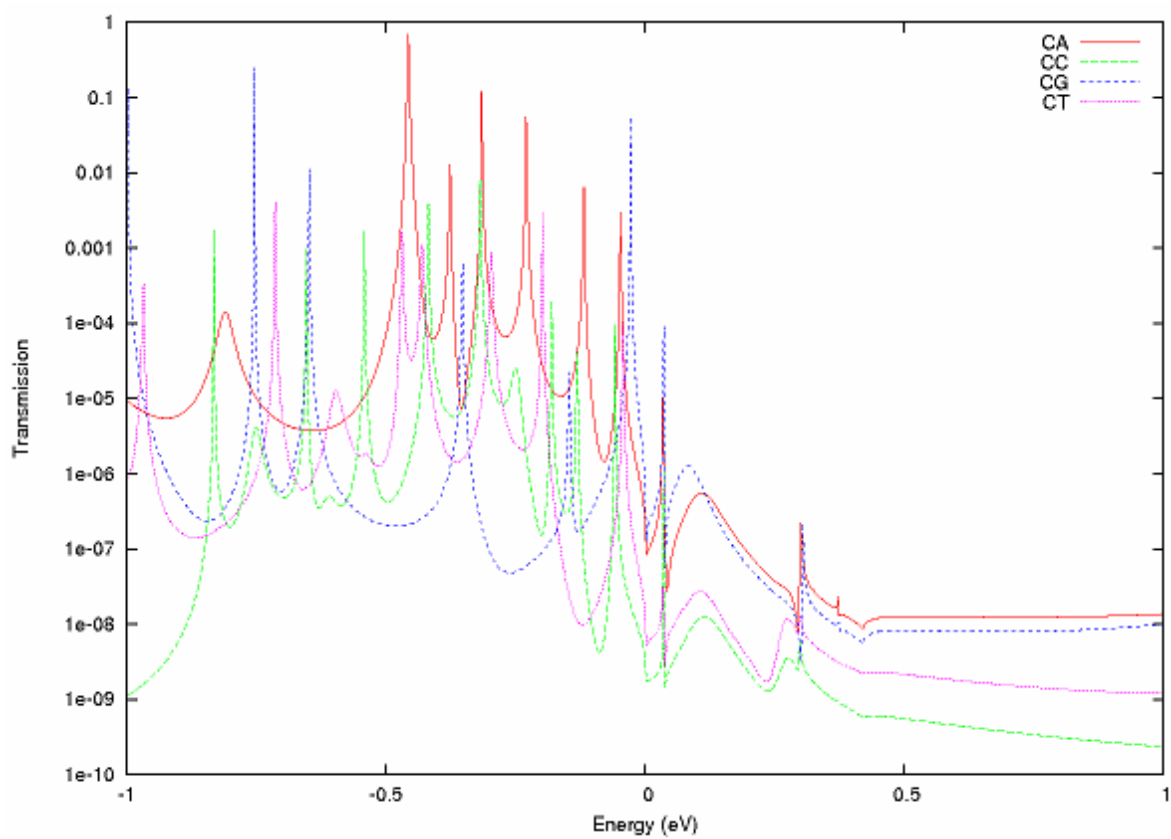

**Figure S1: Plot of transmission functions for C-probe with various target bases for range -1 eV ... +1 eV.**